# A JSON Token-Based Authentication and Access Management Schema for Cloud SaaS Applications

Obinna Ethelbert[†], Faraz Fatemi Moghaddam[*,†], Philipp Wieder[*], Ramin Yahyapour[*,†]

[†]Institute of Informatics, Georg-August-Universität, Göttingen, Germany
[*]Gesellschaft für wissenschaftliche Datenverarbeitung mbH Göttingen (GWDG), Göttingen, Germany
*Email*: c.obinna@stud.uni-goettingen.de, {faraz.fatemi-moghaddam, philipp.wieder, ramin.yahyapour}@gwdg.de

*Abstract*— Cloud computing is significantly reshaping the computing industry built around core concepts such as virtualization, processing power, connectivity and elasticity to store and share IT resources via a broad network. It has emerged as the key technology that unleashes the potency of Big Data, Internet of Things, Mobile and Web Applications, and other related technologies; but it also comes with its challenges – such as governance, security, and privacy. This paper is focused on the security and privacy challenges of cloud computing with specific reference to user authentication and access management for cloud SaaS applications. The suggested model uses a framework that harnesses the stateless and secure nature of JWT for client authentication and session management. Furthermore, authorized access to protected cloud SaaS resources have been efficiently managed. Accordingly, a Policy Match Gate (PMG) component and a Policy Activity Monitor (PAM) component have been introduced. In addition, other subcomponents such as a Policy Validation Unit (PVU) and a Policy Proxy DB (PPDB) have also been established for optimized service delivery. A theoretical analysis of the proposed model portrays a system that is secure, lightweight and highly scalable for improved cloud resource security and management.

## I. Introduction

Cloud computing is a holistic computing construct that encompasses the atomized facets of Information Technology; It is panoptic and embodies not just virtualization, but also dynamic hyper-scaling, elasticity, resource-pooling, isolation and automation [1]. Despite the considerable benefits of cloud-based services, there some significant concerns that have affected the reliability and efficiency of this emerging technology [2].

Understanding the technological backdrop of cloud computing is essential in the analysis of various security and privacy issues in the cloud ecosystem. Such issues range from user authentication, access management, compliance, and recovery, to areas such as data locality, and long term viability. The recent EU referendum that abrogates the US-EU safe Harbor Agreement, to be replaced with a General Data Protection Regulation (GDPR) in 2018 is an issue synonymous with data locality in the cloud ecosystem [3].

Cloud computing has come to stay with all its inherent benefits of increased speed and agility, massive economies of scale, high flexibility, and rapid provisioning [4] or de-provisioning. However, security and privacy issues in the cloud SaaS model has been an area of challenge; in fact, authentication & authorization, confidentiality of data, security of network information and infrastructures, identity management and single-sign on processes [5] are some areas in this regard and which is driving many research work. Many researches are aimed at addressing the challenges of verification and protection of credentials, account hijacking issues, breach of data [6], and simultaneously also, the inherent challenges that arises with increased user access to cloud SaaS resources [7].

Users can be authenticated to access resources via passwords, biometric, token-based or through certificates [8]. But, regardless of the access protocol used, the round-trip security of the access information from the user to the cloud SaaS must be ensured. This requires not only the SaaS provider, but also the app builder, and end user involvement [9]. The secure handling of the access token, and secure user authentication redirect (OpenID connect, JWT, SAML) is a must for the SaaS developer. Also, the authorization process (Oauth2) should be handled through secure HTTPS/TLS transmissions.

Tokens are system generated arbitrary construct that asserts the identity of what it claims to be [10]. Token-based authentication embodies the exchange of client authentication credentials for a server generated authentication token; and for subsequent client requests to access SaaS resources, the tokens are sent as part of the request in the HTTP header to the server. This reuse of the same user access token for accessing protected resources governed by certain policies can be a challenge, especially when a *resource access policy* is updated and the *user access token* is still valid. In fact, this can introduce a Time-Based vulnerability (timing attack) on the protected resources; and with multiple users accessing the resource, the vulnerability index can increase exponentially. Hence, an authentication and authorization model that limits such vulnerabilities and enhances secure resource access have been proposed and evaluated.

## II. JSON Web Token (JWT)

JWT is a standardized tripartite (*Header, Payload and Signature*) token structure that is encoded in a compact JSON serialization format (*using Base64-URL*) consisting of JSON Web Signature (JWS) and JSON Web Encryption (JWE) [10]. Both serializations use different keys for signature and encryption. The JSON Web Key (JWK) and JSON Web Algorithm (JWA) are cryptographically embedded in each JWT. Its seamless compatibility with X.509 key certificate makes it able to carry more information. RFC7519 discusses implementation standards for how JWTs can be signed and encrypted [12]. Tokens can be access tokens, refresh tokens or identity tokens [10]. JWT access tokens can be used for validation of subsequent client request without making frequent calls to the resource server or database. This feature of JWT access tokens can abridge the service latency of OAuth2 [13]. JWS and Public Key Cryptography (PKC) are used for the access token validation. Access tokens can have limited validity periods via embedded expiration time. Access-related claims can also be embedded as part of its payload. While access tokens map user access request to cloud SaaS resources, Identity tokens are useful in Single Sign On (SSO) and identity federation scenarios where users can access different SaaS resource securely without frequent credential provision [6]. However, many SaaS apps in the cloud ecosystem utilize the generic SSO and authorization process (Fig. 1.) [14] for access management regardless of the policies that might govern such SaaS resources.

When SaaS resources are governed by a standardized and well-structured policy database, accurate mapping of authorized users with the resource policy should be well implemented. This should involve verifying and validating that each user request token to the SaaS resource is in tandem with the current policy governing the SaaS resource; and when the policy of the SaaS resource is dynamically updated, the user access token should either reflect the updated policy or be marked invalid.

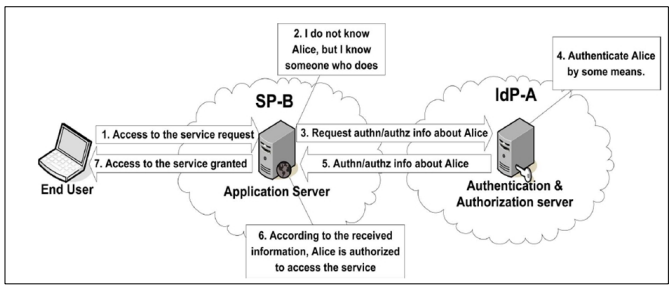

Fig. 1: SSO & Generic Identity Federation

III. RELATED WORKS

Various performed research and manufactured models have been proposed for user authentication and access management. An integrated model for handling identity and access management was proposed in [6]. This incorporates a model for identity management via an identity system and a model for access management authorization via attributes (Attribute Based Access Control - ABAC). This raises the issues of scalability with increased heterogeneous user access to the cloud resource for which an exponentially large number of attributes must be understood and managed for a single resource request call [23]; In addition, because attributes only make sense when they are associated with a user, object, or relation, the practicality of heterogeneous user audits based on attributes in cloud sphere can also be a challenge.

The identity model also utilizes the Security Assertion Markup Language (SAML) and OAuth protocols, whilst the ABAC model utilizes a rule engine via digitally signed predefined XML rules. Nonetheless, whilst SAML stipulates a standard for token creation that is expressive and flexible, it comes at a cost in size and complexity [12]. The token size is further increased by its use of XML and XML Digital Signatures (XML DSIG); not to mention other implementation complexities. This can be a challenge in SaaS implementations requiring simplicity and compactness.

In [15], a fully decentralized identity management framework for personal cloud interoperability was proposed. The framework utilizes JSON Web Token (JWT) and NameCoin as used in block-chain. It proposes a decentralized pattern for sharing public key based on NameCoin (NMC) technology. However, current concerns on technology integration, regulatory issues, cultural trends, *etc.* [16] limit the widespread adoption of the technology.

[17] depicts an S-RBAC model for SaaS systems. The model incorporates an authentication component, access filter server, access control server, a user dynamic constraints server, and a permission management center. These components work concurrently to provide secure access to resources. However, classical scalability issues of the RBAC like permanent role assignment, constraints on time and authorization periods, rule expression based on context, and vulnerability to covert channels [18] still affect the model.

In [19], an integrated architecture for cloud identity and access management was proposed. It involves four components – Cloud Service Provider (CSP), identity management, policy management, Resource Engine and the Access Decision making component. The architecture depicts a workflow for user access to resources either by a CSP, or uploaded by the users themselves. However, a particular SaaS resource may be composed of a plethora of micro-services for which access policy to each micro-service should be unique rather than having a generalist resource access policy. This implies that a proper resource access categorization cum micro service categorization should be well implemented to reflect judicious combination of attributes and roles of both users and resources [23]. In fact, the most challenging issue in this part is to define, manage and map access policies according to capabilities of service provider and requirements of subscribers [24].

Furthermore, a cloud token management system was proposed in [20] for verifying user authorization and data correctness. This is based on a behavior-based token generation process for the tripartite (*upload, download, update*) activities carried out by the user on the cloud resource. It implies that the token management system generates a token for every process/event performed by the user. This raises some concerns over some authorized users' processes with malicious intents that get a valid token from the TMS, which might lead to resource corruption, downtime, and other issues. Therefore, the TMS should move beyond generation and logging, to incorporate resource-driven user access mapping and management.

[21] and [22] describe the application of token based authentication in different scenarios, and specify various token

structures. While structures may differ in design and specifications, none offers a token-based dynamic resource policy management for user accesses via the statelessness and compactness provided by JWT.

## IV. PROPOSED MODEL

The proposed model facilitates user access to protected resources (SaaS resources or micro-services can have sensitive or non-sensitive data. That is, resource categorization based data sensitivity) via secure JWT access tokens.

These accesses are monitored and mapped to be in tandem with the current Resource Access Policy (RAP). Every user request call to the resource goes through a Policy Match gate (PMG) and each event or process is monitored by a Policy Activity Monitor (PAM) which ensures that the user access tokens making the request call have the appropriate RAP, even though the token is still valid. This use of JWT tokens both at client and server side rather than a SAML tokens ensures statelessness and compactness, and thereby reduces overheads which might occur from frequent calls and transmissions of the tokens via HTTP headers.

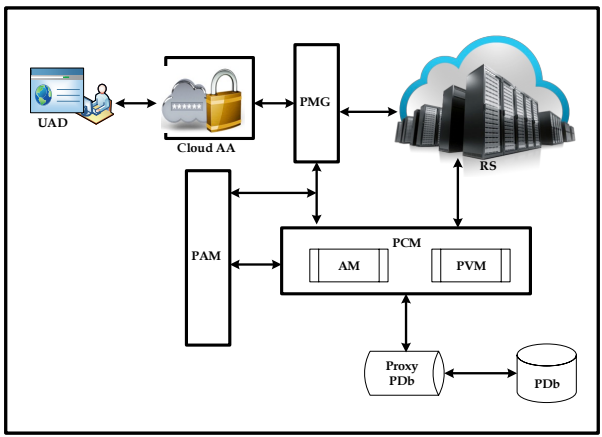

Fig. 2. Proposed model

The proposed model encompasses the following processes:

- User authentication and authorization.
- Authorized user access to protected resources – with valid access token and no update on the RAP.
- Authorized user access to protected resources – with valid access token and an update on the RAP.

### A. Pre-Requisites

These processes assume the following pre-requisites:

o Availability of a well-configured Policy Database.
  o Since implementations might be on a case by case basis, we suggest using standard industry benchmarks (such as: low vulnerability index, scalability and adaptability, ease of integration, XACML formats, *etc*.) for determining the apposite Policy Database configuration. Such policy database can be defined by an enterprise and stored with a cloud provider.

o Cloud based Identity providers (IdPs) and Authorization Servers (AS) are trusted.
  o Using a trusted cloud based IdPs and AS is pertinent for SaaS resources. Similarly, AS and IdPs should be chosen with regards to use cases via standardized industry benchmarks.

o Users are registered users with the IdPs. This is necessary because unregistered users will have to go through a registration phase to obtain their digital identities.

o Re-direct communication is via TLS 1.2 and HTTPS
  o This use of TLS 1.2 and HTTPS is because of the obvious cryptographic advantages of AES and MD5-SHA1 which provides secure end-to-end transport for the tokens.

### B. Components

The model delineates the following components which aid to provide continuous secure access to resources even when the resource access policy is updated and the user access tokens are still valid.

#### 1) Policy Match Gate

This is the connecting and mapping interface between the user access rights and the defined policies in the policy database. Python Web Service Gateway Interface (WSGI) can be used for the implementation of the PMG. The PMG ensures that the user, even while his token is still valid, can only access the protected resource or micro-service specified by the Policy Database (PD). It's the interface between the user POST and GET calls, and the Resource Server's responses.

#### 2) Policy Control Module

This is the policy admin interface that synchronizes calls to the policy DB and policy validation module (PVM). As a control and admin unit, it interfaces with the PAM, to determine when a policy update is required, it handles policy retrievals from the policy database, policy archiving via interface with the PMG, and can be used for auditing user activities based on matched policies by the PMG and its interface with PAM statistics.

#### 3) Policy Validation Module

A sub-component of the PCM; when a resource policy is updated, validity check of the current valid user token and the new resource access policy is required before being forwarded to the PMG. This provides a double authentication feature and ensures that the user is only matched to the allowed resource access permissions. This validation checks involves the dynamic creation of new payloads in the JWT that is being passed in the request headers. This payload is unique and the JWT dynamically signed based on some strong cryptographic algorithms and resides server side waiting for the next valid user request. This is used to assure correctness of access rights for every process request.

#### 4) Proxy Policy DB

This is a high speed DB cache which can be implemented in memory to reduce frequent calls to the Policy DB. This is necessary because a plethora of heterogeneous users can access a particular SaaS resource having the same policy level, therefore, making frequent calls to the Policy DB for multifarious users on the same resource can increase latency even with optimized DB solutions.

*5) Policy Activity Monitor.*

This is the module for statistics and risk management. It ensures user access rights are within allowed resource policy access provisions via periodic checks with PDB and the Resource Server.

For statistics, rather than creating tokens for each user behavior, it monitors and logs user activities, request calls, and other events based on pre-defined benchmarks, user attributes, user roles, and token validity. Also, for risk monitoring, its interface with the PCM triggers policy decisions and policy update request on a SaaS resource based on user activities. Audit trails and accounting can easily be performed as required either user-driven audits, or resource access audits.

### C. Processes

*1) User Authentication & Authorization*

This is the first process required by all intended users of the SaaS resources regardless of the resource categorization. The user/app/device needs to provide their apposite digital identities which can be verified with the trusted IdPs and AS. Successful identity verification creates an access token for the identity. JWT access tokens are unique to each identity and are periodically persisted with a short expiration time at client side for subsequent calls that reuse the access tokens, and also to prevent unauthenticated access.

```
UAD sends RFA via AuthC
RS verifies AuthC
   If AuthC == 200 OK:
       RS generates JWT (See token format below) with GAR
       RS_UAD_response = JWT
       UAD persist JWT for subsequent request calls
   Else:
       RS R-AuthC to AS
       AS verifies AuthC
       If AuthC == 401:
           AS R-UAD to IdP
           IdP verifies AuthC
           If AuthC == 401:
               UAD_registration = UAD-R
```

TABLE I. THE NOTATIONS OF PROPOSED MODEL

| Notation | Descriptions |
|---|---|
| UAD | User, App, or Device |
| AuthC | Authentication Credentials: Username, Password, Location ID, IP Address, etc. |
| 401 | HTTP unauthorized access code |
| RFA | Request for Access |
| RS | Resource Server: Server store for the SaaS resource |
| AS | Authorization Server |
| GAR | Granted Access Rights |
| R- | Redirect- |
| -R | -Required |

HTTP redirects between the AS, RS and IdP are used for the initial authentication and access authorization to protected resources. This utilizes the advantages of OAuth2.0. but because the tokens generated are JWT, it reduces the latency of OAuth 2.0 services for subsequent request calls that require validating the access tokens. In addition, the model proposes using the structured JWT registered claims but with a unique custom ID claim that references the RAP updates to a valid JTI (JWT ID tokens).

*Format:*

Header:
```
{
   "alg": "HSHA256",
   "typ": "JWT"
   "kid": "IdP's JWK set URL"
}
```

Payload:
```
{
   "iss":   "#appbackend",
   "sub":   "#coyname",
   "aud":   "#appfrontend, intended UAD",
   "exp":   "#timeofexpiration(posix format)",
   "nbf":   "#null or #setastartdate",
   "iat":   "#timeofexpiration(posix format)",
   "jti":   "#unique JWT ID",
   "rapID": "#unique identifier for resource access
            policies(rap): It can be a string, number,
            or an array".
}
```

Signature:
```
{
   "Base64-encoded (Header.Payload)" + "private key" + "Algorithm"
}
```

JWT Token =

$$f(\text{base64Encode}) \sum_{n=\alpha,\beta}^{\infty} (header.payload.signature)$$

\* The "rapID" payload is part of the JWT and contains the following data:

**"rapID": ["rap_iat", "rap_Tno", "rap_V", "rap_reqC", "rap_jti"]**

Where:
*rap_iat* = issued timestamp for the updated policy.
*rap_Tno* = total number of policies updated.
*rap_V* = validity check flag for the policy.
*rap_reqC* = required credentials from the user.
*rap_jti* = the unique jti for which this rapID is mapped to.

*2) Authorized User Access with Valid Token & No Update on the RAP*

Here, authenticated and authorized users have access to the protected resource based on their initial AuthC and access privileges. Token persisted at the UAD site carry payload data which contain the GAR privileges. The rapID during this process is set to a value representing the current state of the UAD token, the RAP update level, and the RAP update requirements; and of course, all subsequent UAD call to the resource server are monitored by the PAM for assurance.

*3) Authorized Access with Valid Token & Updates on the RAP*

Because JWT are stateless, persisted at client sites, and are contained in HTTP headers that are used for subsequent request calls to the RS as long as its "exp" payload is still valid, authorized users with malicious intents can have access to resources for which they aren't supposed to. A combination of resource- and user- driven access control mechanism via JWT can be a solution. Harnessing the rapID claim on the JWT that is mapped to a valid JWT assures that users are always within resource policy boundaries.

When the RS updates the RAP for a particular resource while being accessed by an authorized user with a valid token, process-traps are triggered by the PAM requiring the UAD to update its AuthC to be in tandem with the new RAP. This process-trap also set the rapID payload data at the RS server side as this new valid token for the user JTI. Once supplied, the RS redirects to the PMG for proper mapping of supplied credentials; PMG checks if the policy validity is synonymous with the AuthC via the PCM. Once confirmed ok, the UAD is issued a new JWT with the same JTI and the rapID set to the current state of the UAD token, the RAP update level, and the RAP update requirements.

### D. Sample Use-case Schematics

This use case describes a simple scheme for the model's capabilities and components; together with the inter-relationship between them.

- User A requires access to a particular SaaS resource X (SaaSR-X) residing on Resource Server 2 (RS2).
- User A need to authenticate himself via the trusted IdPs and AS. This is represented as (cloud AA)
- Once authenticated, RS2, in tandem with PMG creates a JWT with a JTI and a rapID set that is mapped to User-A.
- Subsequent access to SaaSR-X requires reuse of the created JWT and is checked at the PMG and monitored by the PAM per call if the "exp" payload as configured on the JWT is still valid and there is no RAP update on SaaSR-X.
- User-A decides to access another SaaS resource Y (SaaSR-Y) still residing on Resource Server 2 (RS2) or there is an update on the SaaSR-X.
- PAM notices and triggers a PMG check (process-trap) for userA AuthC and rapID value, if OK, access is granted, else, RS2 request for updated AuthC from User-A.
- User-A supplies the updated AuthC required to access SaaSR-Y or to match the newly updated RAP on SaaSR-X. RS2 redirects to PCM for validation and confirmation, if OK, PCM triggers the PMG for a possible match to the exact GAR of User-A. RS2 uses the PMG matched info, obtained from the rapID to create a new JWT for User-A. User-A valid token is either updated to reflect the new GAR or blacklisted, and a new one issued.

## V. EVALUATION

Theoretical analysis of the model promises the following competitive objectives and security indices.

### A. Security

*MITM:* The model adopts a dual authentication mechanism for each UAD requests at the PMG and at the RS. This ensures concurrency of policy and token per UAD request and assures security of cloud resource. However, when the RAP policy is dynamically updated based on UAD activity that is monitored by the PAM, RS listens for UAD response. While waiting, malicious hackers cannot tamper with the info because nothing has been sent from the RS; but from the PMG, and the rapID which was used by the RS has changed, which renders any token residing at the client side or in transit invalid, thus making the probability of token interception, almost zero.

*Timing Attacks*: An analytical side-channel attack that exploits computational and communication data interactions on crypto-processing systems [25]. Ensuring all received JWTs are duly verified in a lucid and succinct verification contract of allowed crypto "alg", and other payload/header data cum the use of the "kid" makes the verification cut in stone without undergoing unauthorized mutation.

### B. Scalability and Efficiency

Cloud based authentication via trusted IdPs promises capability to meet increased user access demands. Also, the stateless and compact feature of the JWT UAD GAR are highly portable via HTTPS headers without undue latency during peak demands.

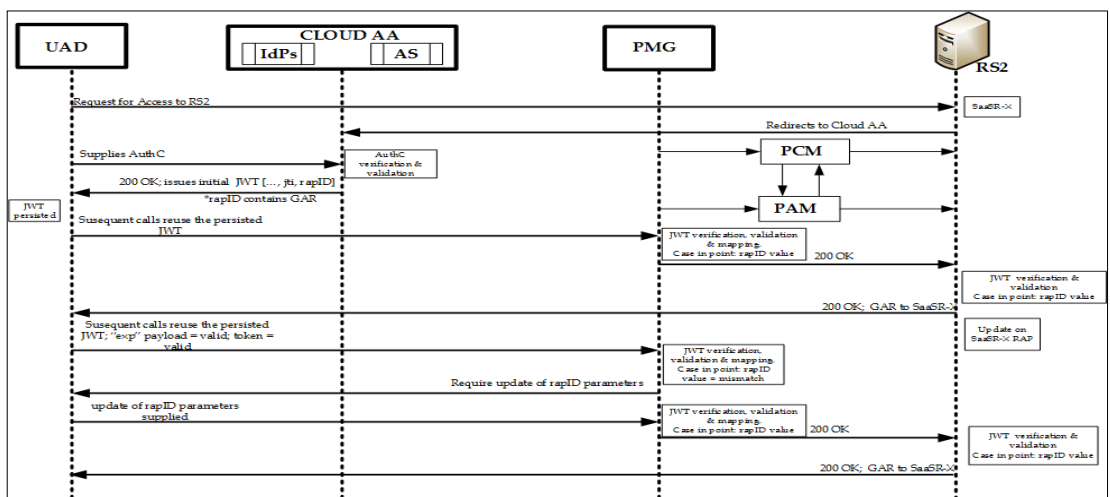

**Fig. 3.** Process Scenario

TABLE II. COMPETITIVE ANALYSIS

| Objectives | Proposed Model | Dancheng | Indu | Sudha | Balamurugan |
|---|---|---|---|---|---|
| **Authentication mechanism** | Token based (JWT) | - | Token based | Token based | Token based |
| **Access control mechanism** | Token based (JWT) | RBAC | ABAC | Token based | Token based |
| **Compact & stateless token structure** | Yes | No | No | No | No |
| **Dual authentication / SSO support** | Yes | No | Yes | No | No |
| **Access control Scalability** | Yes | No | Yes | No | No |

## VI. CONCLUSION

A token-based authentication and access control model was proposed in this paper. The paper leverages on the stateless and compact feature of JWT for authentication and access authorizations for cloud users. There was the succinct introduction into basic cloud computing and security concepts that glides into related research works. JWT was explored with regards to its format, crypto algorithms, and key ids. A proper implementation of the payload data via the introduction of a resource access policy id was seen to enhance needed authorization challenges. Also, components such as PMG, PCM, and PAM were introduced to ensure process authentication per request call via valid tokens from UAD. The simulated tests and practical implementation of the proposed model will be a cause for future research in other to measure performance statistics at the interfaces of each component and possible security lapses.

## ACKNOWLEDGMENT

This research has been partially supported by **CleanSky** project (607584 Grant No.) funded by the **Marie-Curie-Actions** within the **7th Framework Program of the European Union** (**EU FP7**).